# Spin-frustrated pyrochlore chains in the volcanic mineral kamchatkite (KCu$_3$OCl(SO$_4$)$_2$)

**L. M. Volkova[1] and D. V. Marinin[1]**

**Abstract:** Search of new frustrated magnetic systems is of a significant importance for physics studying the condensed matter. The platform for geometric frustration of magnetic systems can be provided by copper oxocentric tetrahedra (OCu$_4$) forming the base of crystalline structures of copper minerals from Tolbachik volcanos in Kamchatka. The present work was devoted to a new frustrated antiferromagnetic - kamchatkite (KCu$_3$OCl(SO$_4$)$_2$). The calculation of the sign and strength of magnetic couplings in KCu$_3$OCl(SO$_4$)$_2$ has been performed on the basis of structural data by the phenomenological crystal chemistry method with taking into account corrections on the Jahn-Teller orbital degeneracy of Cu$^{2+}$. It has been established that kamchatkite (KCu$_3$OCl(SO$_4$)$_2$) contains AFM spin-frustrated chains of the pyrochlore type composed of cone-sharing Cu$_4$ tetrahedra. Strong AFM intrachain and interchain couplings compete with each other. Frustration of magnetic couplings in tetrahedral chains is combined with the presence of electric polarization.

## Introduction

The matter magnetic properties are to a great extent determined by its crystal structure. An impressive versatility of structural features of exhalative copper minerals from fumaroles of the Great Tolbachik Fissure Eruption (GTFE) (Kamchatka Peninsula, Russia) (Vergasova 2012) must promote respective richness and versatility if their magnetic properties. Complexes of oxocentered tetrahedra (OM$_4$) form the base of many GTFE exhalation minerals. Structural studies of these minerals performed by Filatov and Krivovichev with co-workers (St. Petersburg School of Structural Mineralogy and Crystal Chemistry) made a substantial contribution to the development of a separate research direction in advanced structural mineralogy and inorganic crystal chemistry. This new field of crystal chemistry (Krivovichev et al. 1998, 1999, 2001, 2013) is based on cationic tetrahedra (XA$_4$, for example, [OCu$_4$]$^{6+}$), in which the X anion (for example, oxygen (O) – an "extra" oxygen atom not included into acidic radicals) is the central atom, whereas magnetic Cu$^{2+}$ ions are located in vertices. "Extra" oxygen atoms "pull in" cations, thus forming oxocentered tetrahedra with comparatively high strengths of chemical bonds. These anion-centered tetrahedra could couple to each other, this forming isle-like complexes, infinite chains, layers, or frameworks. Magnetic minerals containing oxocentered tetrahedra (OM$_4$) can be not only of scientific, but also of practical interest, for instance, as new frustrated magnetics for spintronics.

The selection of minerals originated from Kamchatka (including kamchatkite) as the objects of study was determined by two factors. The first one was the existence of a specific geometric configuration in the sublattice of magnetic Cu$^{2+}$ ions as Cu$_4$ tetrahedra. It is known that the triangular geometry of tetrahedra faces could serve as the main factor of frustration of magnetic subsystems, if magnetic couplings between nearest neighbors in triangles are of an antiferromagnetic nature and compatible with respect to the force. In this case, a simultaneous energy minimization for all pairwise interactions in triangles composing the tetrahedra is impossible (Figs. 1a, b).

As regards the second factor, the widely spread representations about the nature of magnetic interactions have a crystal chemical aspect, because they evidently indicate the dependence of the interaction strength and spin orientation of magnetic ions on the arrangement of intermediate anions between magnetic ions [Kramers 1934, Goodenough 1955, 1963; Kanamori 1959, Anderson 1963 Vonsovsky 1971]. The dependence of the nearest neighbor interactions on the M-X-M bonding angle is proven and generally accepted. The Cu–O–Cu angle in a regular oxy-centered tetrahedron is equal to

---

L. M. Volkova
volkova@ich.dvo.ru

[1] Institute of Chemistry, Far Eastern Branch of Russian Academy of Science, 159, 100-Let Prosp., Vladivostok, 690022, Russia

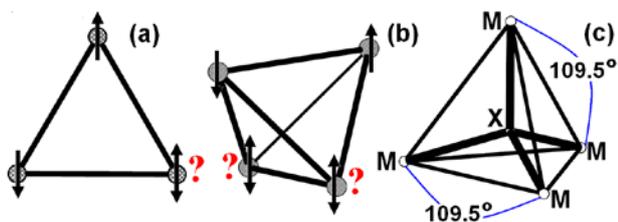

**Fig. 1.** Frustrated "plaquettes": the triangle (a), the tetrahedron (b); $M_i$-X-$M_j$ bonding angle in oxocentered tetrahedra ($OM_4$).

109.5° (Fig. 1c), i.e., exceeds 90°. The latter generates an immediate assumption that magnetic interactions along the tetrahedron edges could be of an AFM character. Therefore, frustration would be a characteristic feature of such a tetrahedron. The shift of the centering ion inside the tetrahedron from its center or the tetrahedron distortion must be significant in order to induce spin reorientation of the AFM→FM type. On the other hand, the centering oxygen atom is clamped in the tetrahedron, since a broad variation of short bonds (not axial Cu–O ones) is impossible. In empty tetrahedra, the magnetic couplings along their edges are significantly weaker, and, besides, the intermediate oxygen ions are capable to varying their positions and reorienting spins of magnetic $Cu^{2+}$ ions.

At present, rather active research activities are concerned with the three-dimensional (3D) network of corner-sharing tetrahedra in geometrically frustrated antiferromagnets $A_2B_2O_7$ of the pyrochlore type ($NaCa(Nb_2O_6)F$) (Gaertner 1930). The crystal structure of these compounds has a centrosymmetric space group Fd3m (N227). In the rare–earth pyrochlore oxides of the formula $R_2M_2O_7$, the trivalent magnetic rare–earth $R^{3+}$ ions (e.g., R = Dy and Ho; M = Ti is nonmagnetic) (Farmer et al. 2014) reside on a three-dimensional (3D) pyrochlore lattice of corner-sharing ($OR_4$) tetrahedra. The magnetic subsystem of such compounds built from tetrahedral $R_4$ blocks is strongly frustrated and, in some cases, completely prevents the formation of the long-range order until realization of exotic states of the "spin ice" or "spin liquid" types (Balents 2010; Bramwell et al. 2001; Greedan 2001; Harris et al. 1997, 1998; Moessner et al. 2006; Morris et al. 2008; Ramirez et al. 1999; Reimers 1992; Reimers et al. 1991; Sosin et al. 2005).

From the 3D pyrochlore lattice, one can separate (cut) chains of corner-sharing tetrahedra – segments of the pyrochlore structure. Such single chains of ($OCu_4$) tetrahedra are contained in the crystal structure of the known quasi-one-dimensional antiferromagentic $Cu_3Mo_2O_9$ (Steineret al. 1997; Reichelt et al. 2005). However, according to (Hamasaki et al. 2008; Hase et al. 2015; Kuroe, Hamasaki et al. 2010 and 2011; Matsumoto et al. 2012, Naruse et al. 2015), the magnetic structure of $Cu_3Mo_2O_9$ does not fully coincide with the crystal structure of the sublattice of magnetic $Cu^{2+}$ ions. The magnetic structure of $Cu_3Mo_2O_9$ comprises a spin-1/2 frustrated antiferromagnet consisting of AFM linear chains and AFM dimers stretched along the $c$ axis. Determination of the magnetic structure of $Cu_3Mo_2O_9$ in neutron powder diffraction experiments (Matsumoto et al. 2012, Hase et al. 2015) indicates to the existence of a partially disordered state explained by the effect of magnetic frustration on the magnetic structure.

The compound $Cu_3Mo_2O_9$ is characterized by weak ferromagnetism, electric polarization (Kuroe, Kino et al. 2011), and many other interesting properties. In (Hamasaki et al. 2008) weak ferromagnetic moments on the $ac$ plane are explained by the Dzyaloshinskii–Moriya interaction in the Cu linear chains. As was shown in (Kuroe, Hamasaki et al. 2011), geometrical magnetic frustration served as the origin of a nontrivial spin configuration that breaks the spatial inversion symmetry and allowed the distorted tetrahedral spin system manifestation of the multiferroic behavior without any magnetic superlattice formation. Studies of magnetic state of the geometrically frustrated quasi-one-dimensional spin system $Cu_3Mo_2O_9$ by thermal conductivity enabled one to assume (Naruse et al. 2015) the existence of a novel field-induced spin state discussed in terms of the possible spin-chirality ordering in a frustrated Mott insulator.

In the present paper we will show the magnetic structure of the noncentrosymmetric mineral kamchatkite ($KCu_3OCl(SO_4)_2$) (Krivovichev et al. 2013) from the point of crystal chemistry. The crystal structure of the sublattice of magnetic $Cu^{2+}$ ions in $KCu_3OCl(SO_4)_2$ contains single pyrochlore chains of ($OCu_4$) tetrahedra. Such chains are also present in $Cu_3Mo_2O_9$. To determine the structure of the magnetic subsystem of this mineral, we calculated the characteristics (sign and strength) of $J_{ij}$ magnetic interactions not only within low-dimension fragments, but also between them at long distances based on the data on the chains ideal crystal structure.

We believe that the frustrated antiferromagnetic systems of this type can be of interest not only for theoretical, but also for experimental studies of their magnetic properties and real magnetic structures, whose formation can be contributed, aside from structural data, by many other physical factors.

**Method of calculation**

To determine the characteristics of magnetic interactions (type of the magnetic moments ordering and strength of magnetic coupling) in minerals kamchatkite $KCu_3OCl(SO_4)_2$ (Varaksina et al. 1990), we used the earlier developed phenomenological method (named the "crystal chemistry method") and the program "MagInter" created on its basis (Volkova and Polyshchuk 2005, 2009; Volkova 2009). In this method, three well-known concepts about the nature of magnetic interactions are used. First, it was the Kramers's idea (1934), according to which in exchange couplings between magnetic ions separated by one or several diamagnetic groups, the electrons of nonmagnetic ions play a considerable role.

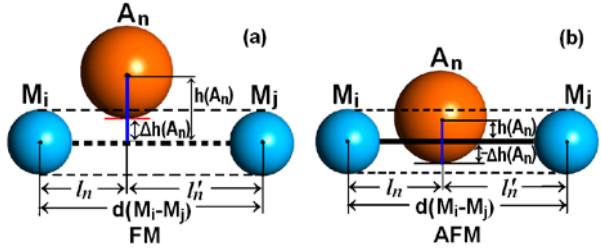

**Fig. 2** A schematic representation of the intermediate A$n$ ion arrangement in the local space between magnetic ions M$i$ and M$j$ in cases where the A$n$ ion initiates the emerging of the ferromagnetic (**a**) and antiferromagnetic (**b**) interactions. $h(An)$, $ln$, $ln'$, and $d(Mi–Mj)$ are the parameters determining the sign and strength of magnetic interactions $Jn$.

Second, we used the Goodenough–Kanamori–Anderson's model (Goodenough 1955, 1963; Kanamori 1959; Anderson 1963), in which crystal chemical aspect points clearly to the dependence of strength interaction and the type of orientation of spins of magnetic ions on the arrangement intermediate anions. Third, we used the polar Shubin–Vonsovsky's model (Vonsovsky 1971), by consideration of magnetic interactions we took into account not only anions, which are valence bound to the magnetic ions, but also all the intermediate negatively or positively ionized atoms, except cations of metals without unpaired electrons.

The method enables one to determine the sign (type) and strength of magnetic couplings on the basis of structural data. According to this method, a coupling between magnetic ions $M_i$ and $M_j$ emerges in the moment of crossing the boundary between them by an intermediate ion $A_n$ with the overlapping value of ~0.1 Å. The area of the limited space (local space) between the $M_i$ and $M_j$ ions along the bond line is defined as a cylinder, whose radius is equal to these ions radii. The strength of magnetic couplings and the type of magnetic moments ordering in insulators are determined mainly by the geometrical position and the size of intermediate $A_n$ ions in the local space between two magnetic ions $M_i$ and $M_j$. The positions of intermediate ions $A_n$ in the local space are determined by the distance $h(A_n)$ from the center of the $A_n$ ion up to the bond line $M_i$-$M_j$ and the degree of the ion displacement to one of the magnetic ions expressed as a ratio $(l_n'/l_n)$ of the lengths $l_n$ and $l_n'$ ($l_n \leq l_n'$; $l_n' = d(M_i - M_j) - l_n$) produced by the bond line $M_i$-$M_j$ division by a perpendicular made from the ion center (Fig. 2).

The intermediate $A_n$ ions will tend to orient magnetic moments of $M_i$ and $M_j$ ions and make their contributions $j_n$ into the emergence of antiferromagnetic (AFM) or ferromagnetic (FM) components of the magnetic interaction in dependence on the degree of overlapping of the local space between magnetic ions ($\Delta h(A_n)$), the asymmetry ($l_n'/l_n$) of position relatively to the middle of the $M_i$-$M_j$ bond line, and the distance between magnetic ions ($M_i$-$M_j$).

Among the above parameters, only the degree of space overlapping between the magnetic ions $M_i$ and $M_j$ ($\Delta h(A_n) = h(A_n) - r_{A_n}$) equal to the difference between the distance $h(A_n)$ from the center of $A_n$ ion up to the bond line $M_i$-$M_j$ and the radius ($r_{A_n}$) of the $A_n$ ion determined the sign of magnetic interaction. If $\Delta h(A_n) < 0$, the $A_n$ ion overlaps (by $|\Delta h|$) the bond line $M_i$-$M_j$ and initiates the emerging contribution into the AFM-component of magnetic interaction. If $\Delta h(A_n) > 0$, there remains a gap (the gap width $\Delta h$) between the bond line and the $A_n$ ion, and this ion initiates a contribution to the FM-component of magnetic interaction. The sign and strength of the magnetic coupling $J_{ij}$ are determined by the sum of the above contributions:

$$J_{ij} = \sum_n j_n$$

The $J_{ij}$ value is expressed in Å$^{-1}$ units. If $J_{ij} < 0$, the type of $M_i$ and $M_j$ ions magnetic ordering is AFM and, in opposite, if $J_{ij} > 0$, the ordering type is FM.

The format of the initial data for the "MagInter" program (crystallographic parameters, atom coordinates) is in compliance with the cif-file in the Inorganic Crystal Structure Database (ICSD) (FIZ Karlsruhe, Germany). The room-temperature structural data KCu$_3$OCl(SO$_4$)$_2$ (Varaksina et al. 1990) (ICSD-66309) and ionic radii of Shannon (1976)) were used for calculations. Radii for $^{IV}$Cu$^{2+}$, $^{VI}$O$^{2-}$, $^{VI}$Cl$^-$ and $^{IV}$S$^{6+}$ are equal 0.57Å, 1.40 Å, 1.81Å and 0.12 Å, respectively.

**Taking into account the specifics of volcanic minerals at magnetic couplings parameters calculation**

The minerals under examination belong to the specific class of magnetic substances, for which two factors (the presence of Cu$^{2+}$ ions with orbital degeneracy (the so-called Jahn–Teller ions) and geometric frustration of oxocentric copper tetrahedra (OCu$_4$)) to a great extent determine their magnetic structure and properties.

*Corrections on orbital degeneracy for Jahn-Teller Cu$^{2+}$ ions*

In the crystal structure, the Jahn–Teller effect yields a significant distortion in coordination of Cu$^{2+}$ ions resulting in the fact that the lengths of axial bonds in copper octahedra exceed those of equatorial bonds, until to the emergence of configurations in the forms of stretched octahedron (4+2), square pyramid (4+1), or flat of square. In many cases, the magnetic structure is characterized with an anomalously strong magnetic anisotropy because of orbital degeneracy of $Cu^{2+}$ ions (Kugel' and Khomskii 1973, 1982; Oleś et al. 2006). The most representative example here is the thoroughly studied compound $KCuF_3$ (Towler et al. 1995; Yamada and Kato 1994; Paolasini et al. 2002), which, while preserving an almost cubic crystal lattice, is characterized by quasi-one-dimensional magnetic properties. The neutron-scattering measurements show (Hutchings et al. 1969; Satija et al. 1980) that $KCuF_3$ comprises a one-dimensional antiferromagnetic. There exist a very strong antiferromagnetic interaction ($J_c$ = 17.5 meV) between spins in the chain along the (001) direction and a very weak ferromagnetic one between chains ($J_a$ = -0.2 meV) ($J_a/J_c$ = 0.01).

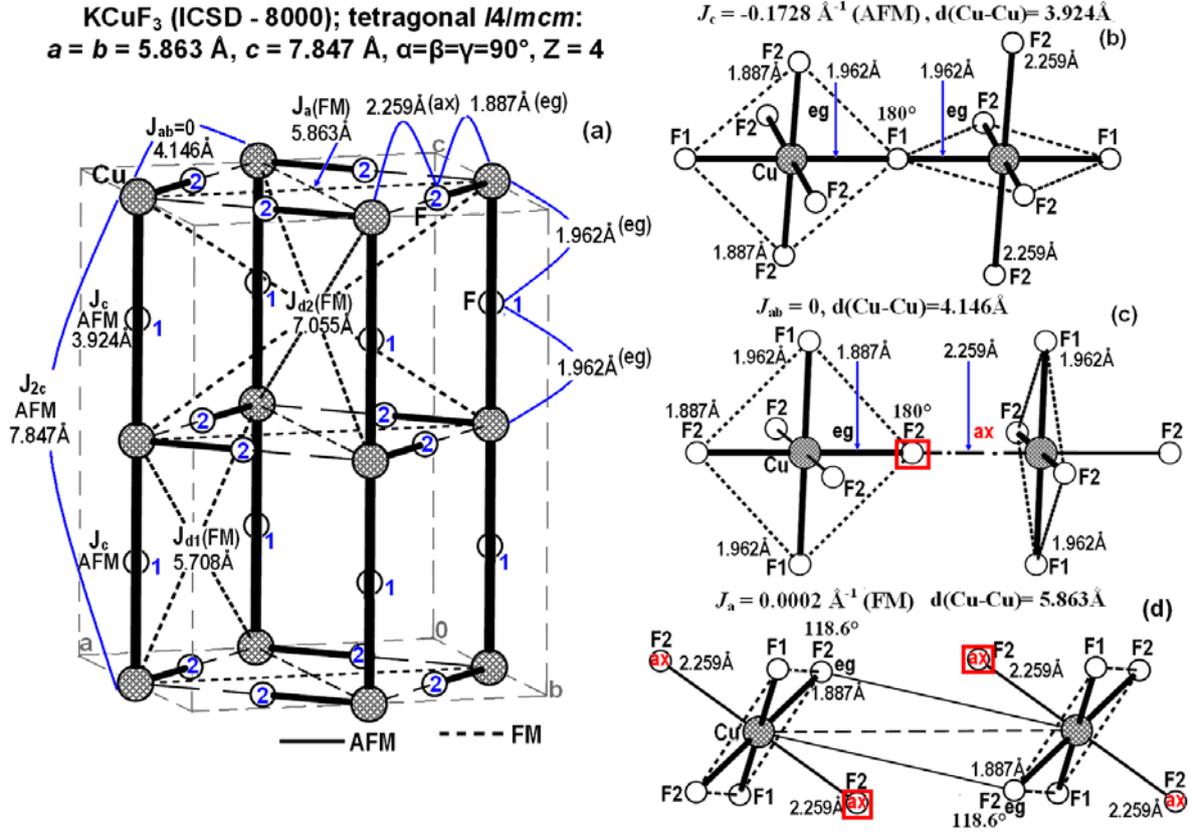

**Fig. 3** The sublattice of magnetic $Cu^{2+}$ ions and the $J_n$ coupling in $KCuF_3$ (a). The arrangement of intermediate ions in the space of AFM $J_c$ (b), $J_{ab}$ (c), and $J_a$ (d) interactions. In this and other figures, the thickness of lines shows the strength of the $J_n$ coupling. AFM and FM couplings are indicated by solid and dashed lines, respectively.

However, in the case of orbital degeneracy, the use of the well-known Goodenough–Kanamori–Anderson rules (Goodenough 1955, 1963; Kanamori 1959; Anderson 1963) does not attaining the similarity between the parameters of exchange interactions and respective experimental data. The crystal chemistry method we developed is also based on the Goodenough–Kanamori–Anderson rules. In order to determine which corrections should be made at this method use for calculations of magnetic interactions of Jahn–Teller ions on the basis of structural data, we examined the compound $KCuF_3$ (Tanaka et al. 1979) as well (Fig. 3). A stretched octahedron (4+2) serves as a coordination polyhedron of $Cu^{2+}$ ions in $KCuF_3$. In this octahedron, four shortened equatorial bonds (two Cu-F2$^{eg}$ bonds and two Cu-F1$^{eg}$ bonds) are equal to 1.887 Å and 1.962 Å, respectively, whereas two axial bonds (Cu-F2$^{ax}$) are elongated until the length of 2.259Å.

According to our calculations, a very strong AFM magnetic interaction ($J_c$) indeed takes place between spins in linear chains along the $c$ axis (Figs. 3ab) ($J_c$ = -0.1728 Å$^{-1}$; d(Cu–Cu) = 3.924 Å). However, it turns out that, unlike the experiment (Hutchings et al. 1969, Satija et al. 1980), a strong AFM $J_{ab}$ coupling ($J_{ab}/J_c$ = 0.91; d(Cu–Cu) = 4.146 Å) also exists between these chains in the $ab$ plane, if one takes into account AFM contributions from all intermediate fluorine ions (Fig. 3c) that entered the local space of the $Cu^{2+}$–$Cu^{2+}$ interaction. It is possible to attain the similarity to the neutron-scattering measurements data only in the case, if one excludes from calculations contributions from intermediate F ions involved into a direct axial Cu–$F^{ax}$ bond with at least one of two $Cu^{2+}$ ions participating in the interaction. As a result, the coupling strength $J_{ab}$ (d(Cu–Cu) = 4.146 Å; Cu–F2…Cu: 1.887Å$^{(eg)}$ and 2.259 Å$^{(ax)}$) in the *ab* plane will be equal to zero. Other interchain couplings ($J_{d1}$ ($J_{d1}/J_c$= -0.010, d(Cu–Cu) = 5.708 Å), $J_a$ ($J_a/J_c$= -0.001, d(Cu–Cu) = 5.863 Å) (Fig. 2d) and $J_{d2}$ ($J_{d2}/J_c$= -0.021, d(Cu–Cu) = 7.055 Å)) are weak ferromagnetic ones. Besides, our calculations demonstrate that there exists a competition between the AFM nearest-neighbor $J_c$ and AFM next-nearest-neighbor $J_{2c}$ ($J_{2c}/J_c$ = 0.19; d(Cu–Cu) = 7.847 Å) intrachain couplings in linear chains along the c axis.

*The scaling factors Kn for translating the value in per angstrom into meV*

The crystal chemistry method we use was created to search for magnetic compounds with a specified magnetic structure based on the data on the compound crystal structure. In order to conclude on the type of the magnetic structure, it is sufficient to know a sign (spin orientation – AFM or FM) and relative values of the forces of magnetic interactions between magnetic ions, as well as specific geometric configurations in their sublattice, on which the frustration of magnetic interactions is possible. Unfortunately, our method allows determining the strength of magnetic interactions that could be possible in the absence of competition – in the absence of barriers to their simultaneous existence from the side of geometric configurations in the magnetic ions sublattice. However, in the frustrated fragments, it is possible to observe an order-of-magnitude difference between the theory and the experiment in the course of determination of the strength of magnetic couplings. To estimate the strength of magnetic couplings with taking into account the frustration and to translate the value $J_n$ in per angstrom (Å$^{-1}$) into the energy units (meV) more conventional for experimenters, it is necessary to select a magnetic fragment similar in crystal structure and chemical composition, which was studied experimentally, to calculate parameters of magnetic couplings by the crystal chemistry method based on the structural data, and to determine the coefficients ($K_n$) of the relationship between theoretical and experimental data for each individual coupling. It is worth emphasizing that different conversion factors ($K_n$) were used for different $J_n$ couplings, since, simultaneously with translating the $J_n$ values in Å$^{-1}$ into those in meV, a correction of the force of the magnetic coupling on the existence of its competition with other couplings in the structure was introduced.

We calculated, using our crystal chemistry method, the parameters of magnetic couplings in two similar single chains of corner-sharing oxocentered tetrahedra ($OCu_4$) in the frustrated quasi-one-dimensional antiferromagnetic ($Cu_3Mo_2O_9$) (Reichelt et al. 2005) and kamchatkite ($KCu_3OCl(SO_4)_2$) (Varaksina et al. 1990) (Figs. 4, 5, Table 1). Thereafter, we calculated the data of calculations of the parameters of magnetic couplings in the $Cu_3Mo_2O_9$ chain by the crystal chemistry method with the experimental data (Matsumoto et al. 2012). It can be concluded that the intermediate ions, whose bond has the Jahn–Teller stretching, do not contribute to the magnetic coupling, as in the case of $KCuF_3$.

Besides, based on the above data, we determined the scaling factors $K$n for each individual coupling in single chains of $Cu_4$ tetrahedra.

Table 1 shows the crystallographic characteristics and parameters of magnetic couplings ($J$n$^{str}$) in Å$^{-1}$ and meV ($J$n = $K$n×$J$n$^{str}$) calculated on the basis of structural data and respective distances between magnetic ions $Cu^{2+}$ in $Cu_3Mo_2O_9$ and $KCu_3OCl(SO_4)_2$. Besides, for all intermediate X ions that entered the local space of $Cu^{2+}$–$Cu^{2+}$ interaction and provided the maximal contributions ($j^{max}$) into AFM or FM components of these $J$n couplings, the degree of overlapping of the local space between magnetic ions $\Delta h(X)$, the asymmetry $l_n'/l_n$ of the position relatively to the middle of the $Cu_i$–$Cu_j$ bond line, and the $Cu_i$–X–$Cu_j$ angle are presented. However, the contributions $j(X^{ax})$ from intermediate ions having direct axial Cu–$X^{ax}$ with at least one of two $Cu^{2+}$ ions were not taken into account. The arrangement of intermediate ions in the local space of $J$1–$J$4 couplings in tetrahedra chains in $KCu_3OCl(SO_4)_2$ are shown in Fig. 5. Similar positions of intermediate ions characterize respective couplings in $Cu_3Mo_2O_9$.

We have not considered the random disorder (for instance, oxygen or cation vacancies, nonmagnetic impurities in positions of magnetic ions etc.), to which the parameters of magnetic interactions could be extremely sensitive.

**Results and Discussion**

Kamchatkite ($KCu_3OCl(SO_4)_2$) (Krivovichev et al. 2013, Varaksina et al. 1990) crystallizes in the noncentrosymmetric orthorhombic $Pna2_1$ system. Copper ions occupy three crystallographically independent sites (Cu1, Cu2, and Cu3) and have the typical [4+2] Jahn-Teller distortion. In Cu1 and Cu2 ions octaherda, equatorial positions are occupied by three oxygen ions and one chlorine atom at short (d(Cu–O) = 1.92–2.08 Å and d(Cu–Cl) = 2.38–2.40 Å) distances and two oxygen ions the octahedron apical positions at long (d(Cu–O) = 2.31–2.37 Å) distances. In the Cu3 octahedron, equatorial and apical positions are occupied only by oxygen ions at short (d(Cu–O) = 1.86–2.043 Å) and long (d(Cu–O) = 2.37–2.43 Å) distances, respectively.

**Table 1** The crystallographic characteristics and parameters of magnetic couplings $J_n$ calculated on the basis of structural data and respective distances between magnetc ions $Cu^{2+}$ in single chains of corner-sharing $Cu_4$ tetrahedra in $Cu_3Mo_2O_9$ and $KCu_3OCl(SO_4)_2$.

| Crystallographic and magnetic parameters | $Cu_3Mo_2O_9$ (Reichelt et al. 2005) (Data for ICSD - 154263) Space group $Pnma$ (N62): $a$ = 7.685 Å, $b$ = 6.872 Å, $c$ = 14.642Å $\alpha$ =90º, $\beta$ = 90º, $\gamma$ = 90º, Z = 4 Method[a] – XDS; R-value[b] = 0.021 | $KCu_3OCl(SO_4)_2$ (Varaksina et al. 1990) (Data for ICSD - 66309) Space group $Pna2_1$ (N33): $a$ = 9.741 Å, $b$ = 12.858 Å, $c$=7.001 Å $\alpha$ =90º, $\beta$ = 90º, $\gamma$ = 90º, Z = 4 Method[a] – XDS; R-value[b] = 0.055 |
|---|---|---|
| Bond $1_1$ | Cu1-Cu2 | Cu1-Cu3 |
| d(Cu-Cu) (Å) | 2.950 | 2.981 |
| $j(X)^c$ (Å$^{-1}$) | $j$(O1$^{eg\,(j)}$): -0.0469 (AFM) | $j$(O9$^{eg}$): -0.0318 (AFM) |
| ($\Delta h(X)^d$ Å, $l_n'/l_n^e$, CuXCu$^f$) | (-0.203, 1.07, 101.87°) | (-0.141, 1.05, 99.61°) |
| $j(X)^c$ (Å$^{-1}$) | $j$(O5$^{ax\,(k)}$): 0.0267 (FM) | $j$(O8$^{ax}$): 0.0908 (FM) |
| ($\Delta h(X)^d$ Å, $l_n'/l_n^e$, CuXCu$^f$) | (0.113, 1.28, 86.13°) | (0.403, 1.07, 79.15°) |
| $J1_1^{str\,(g)}$ (Å$^{-1}$) | -0.0469 (AFM) | -0.0318 (AFM) |
| $J1_1^{(h)}$ (meV) | -3.06 (AFM) (Matsumoto et al. 2012) | -2.07 (AFM) |
| $K1^{(i)}$ | 65.24 | |
| Bond $1_1'$ | | Cu1-Cu3 |
| d(Cu-Cu) (Å) | | 2.997 |
| $j(X)^c$ (Å$^{-1}$) | | $j$(O9$^{eg}$): -0.0547 |
| ($\Delta h(X)^d$ Å, $l_n'/l_n^e$, CuXCu$^f$) | | (-0.245, 1.09, 104.75°) |
| $J1_1'^{\,str\,(g)}$ (Å$^{-1}$) | | -0.0547 (AFM) |
| $J1_1^{(h)}$ (meV) | | -3.57 (AFM) |
| Bond $1_2$ | Cu2-Cu2 | Cu1-Cu1 |
| d(Cu-Cu) (Å) | 5.901 | 5.978 |
| $j(X)^{(c)}$ (Å$^{-1}$) | $j$(Cu1): -0.0327 (AFM) | $j$(Cu3): -0.0297 (AFM) |
| ($\Delta h(X)^d$ Å, $l_n'/l_n^e$, CuXCu$^f$) | (-0.570, 1.00, 180°) | (-0.531, 1.01, 179°) |
| $j(X)^c$ (Å$^{-1}$) | $j$(O1$^{eg}$): -0.0020×2 (AFM) | $j$(O9$^{eg}$): -0.0026 (AFM) |
| ($\Delta h(X)^d$ Å, $l_n'/l_n^e$, CuXCu$^f$) | (-0.203, 2.87, 126.61°) | (-0.263, 2.85, 129.4°) |
| $j(X)^c$ (Å$^{-1}$) | – | $j$(O9$^{eg}$): -0.0011 (AFM) |
| ($\Delta h(X)^d$ Å, $l_n'/l_n^e$, CuXCu$^f$) | – | (-0.124, 3.15, 122.7°) |
| $J1_2^{str\,(g)}$ (Å$^{-1}$) | -0.0367 (AFM) | -0.0334 (AFM) |
| $J1_2^{str}/J1_1^{str}$ | 0.78 | 1.05 |
| $J1_2^{str}/J1_1'^{\,str}$ | – | 0.61 |
| Bond $2_1$ | Cu1-Cu3 | Cu2-Cu3 |
| d(Cu-Cu) (Å) | 2.997 | 3.040 |
| $j(X)^c$ (Å$^{-1}$) | $j$(O1$^{eg}$): -0.0503 (AFM) | $j$(O9$^{eg}$): -0.0605 |
| ($\Delta h(X)^d$ Å, $l_n'/l_n^e$, CuXCu$^f$) | (-0.225, 1.08, 103.76°) | (-0.279, 1.05, 107.19°) |
| $j(X)^c$ (Å$^{-1}$): | $j$(O4$^{eg}$): 0.0152 (FM) | $j$(O1$^{ax}$): 0.0385 |
| ($\Delta h(X)^d$ Å, $l_n'/l_n^e$, CuXCu$^f$) | (0.067, 1.20, 90.98°) | (0.170, 1.36, 87.51°) |
| $J2_1^{str\,(g)}$ (Å$^{-1}$) | -0.0351 (AFM) | -0.0605 (AFM) |
| $J2_1^{(h)}$ (meV) | -3.06 (AFM) (Matsumoto et al. 2012) | -5.27 (AFM) |
| $K2^{(i)}$ | 87.18 | |
| Bond $2_1'$ | | Cu2-Cu3 |
| d(Cu-Cu) (Å) | | 3.014 |
| $j(X)^{(c)}$ (Å$^{-1}$) | | $j$(O9$^{eg}$): -0.0366 |
| ($\Delta h(X)^d$ Å, $l_n'/l_n^e$, CuXCu$^f$) | | (-0.166, 1.05, 101.35°) |
| $j(X)^{(c)}$ (Å$^{-1}$): | | $j$(O4$^{ax}$): 0.0457 |
| ($\Delta h(X)^d$ Å, $l_n'/l_n^e$, CuXCu$^f$) | | (0.198, 1.37, 85.99°) |
| $J2_1'^{\,str\,(g)}$ (Å$^{-1}$) | | -0.0366 (AFM) |
| $J2_1'^{\,(h)}$ (meV) | | -3.19 (AFM) |
| Bond $2_2$ | Cu3-Cu3 | Cu2-Cu2 |
| d(Cu-Cu) (Å) | 5.994 | 6.054 |
| $j(X)^c$ (Å$^{-1}$) | $j$(Cu1): -0.0317 (AFM) | $j$(Cu3): -0.0294 (AFM) |
| ($\Delta h(X)^d$ Å, $l_n'/l_n^e$, CuXCu$^f$) | (-0.570, 1.0, 180°) | (-0.539, 1.01, 178.83°) |
| $j(X)^c$ (Å$^{-1}$) | $j$(O1$^{eg}$): -0.0022×2 (AFM) | $j$(O9$^{eg}$): -0.0027 (AFM) |
| ($\Delta h(X)^d$ Å, $l_n'/l_n^e$, CuXCu$^f$) | (-0.225, 2.85, 128.10°) | (-0.288, 2.88, 130.65°) |
| $j(X)^c$ (Å$^{-1}$) | | $j$(O9$^{eg}$): -0.0014 (AFM) |
| ($\Delta h(X)^d$ Å, $l_n'/l_n^e$, CuXCu$^f$) | | (-0.157, 3.15, 124.46°) |
| $J2_2^{str\,(g)}$ (Å$^{-1}$) | -0.0361 (AFM) | -0.0335 (AFM) |
| $J2_2^{str}/J2_1^{str}$ | 1.03 | 0.55 |
| $J2_2^{str}/J2_1'^{\,str}$ | – | 0.92 |

| Crystallographic and magnetic parameters | $Cu_3Mo_2O_9$ | $KCu_3OCl(SO_4)_2$ |
|---|---|---|
| Bond 3 | Cu2-Cu3 | Cu1-Cu2 |
| d(Cu-Cu) (Å) | 3.169 | 3.183 |
| $j(X)^c$ (Å$^{-1}$) | $j(O1^{eg})$: -0.0545 (AFM) | $j(O9^{eg})$: -0.0641 (AFM) |
| ($\Delta h(X)^d$ Å, $l_n'/l_n^e$, $CuXCu^f$) | (-0.274, 1.0, 109.18°) | (-0.325, 1.01, 111.90°) |
| $j(X)^c$ (Å$^{-1}$) | $j(O3^{ax})$: 0.0730 (FM) | $j(Cl^{eg})$: -0.0054 (AFM) |
| ($\Delta h(X)^d$ Å, $l_n'/l_n^e$, $CuXCu^f$) | (0.340, 1.48, 83.67°) | (-0.027, 1.01, 83.51°) |
| $J3^{str\,(g)}$ (Å$^{-1}$) | -0.0545 (AFM) | -0.0695 (AFM) |
| $J3^{(h)}$ (meV) | -5.7 (AFM) (Matsumoto et al. 2012) | -7.27 (AFM) |
| $K3^{(i)}$ | 104.59 | |
| Bond $4_1$ | Cu1-Cu1 | Cu3-Cu3 |
| d(Cu-Cu) (Å) | 3.436 | 3.501 |
| $j(X)^c$ (Å$^{-1}$) | $j(O1^{eg})$: -0.1164 (AFM) | $j(O9^{eg})$: -0.1004 (AFM) |
| ($\Delta h(X)^d$ Å, $l_n'/l_n^e$, $CuXCu^f$) | (-0.687, 1.0, 134.92)° | (-0.613, 1.08, 131.55°) |
| $J4_1^{str\,(g)}$ (Å$^{-1}$) | -0.1164 (AFM) | -0.1004 (AFM) |
| $J4_1^{(h)}$ (meV) | -6.5 (AFM) (Matsumoto et al. 2012) | -5.61 (AFM) |
| $K4^{(i)}$ | 55.84 | |
| Bond $4_2$ | Cu1-Cu1 | Cu3-Cu3 |
| d(Cu-Cu) (Å) | 6.872 | 7.001 |
| $j(X)^{(c)}$ (Å$^{-1}$): | $j(Cu3)$: -0.0241 (AFM) | $j(Cu3)$: -0.0219 (AFM) |
| ($\Delta h(X)^d$ Å, $l_n'/l_n^e$, $CuXCu^f$) | (-0.570, 1.0, 180°) | (-0.537, 1.00, 178.92°) |
| $j(X)^c$ (Å$^{-1}$): | $j(O1^{eg})$: -0.0048×2 (AFM) | $j(O9^{eg})$: -0.0043 (AFM) |
| ($\Delta h(X)^d$ Å, $l_n'/l_n^e$, $CuXCu^f$) | (-0.687, 3.00, 149.58°) | (-0.599, 2.87, 147.36°) |
| $j(X)^c$ (Å$^{-1}$): | – | $j(O9^{eg})$: -0.0041 (AFM) |
| ($\Delta h(X)^d$ Å, $l_n'/l_n^e$, $CuXCu^f$) | – | (-0.626, 3.14, 147.13°) |
| $J4_2^{str\,(g)}$ (Å$^{-1}$) | -0.0337 (AFM) | -0.0303 (AFM) |
| $J4_2^{str}/J4_1^{str}$ | 0.29 | 0.30 |

[a] XDS – X-ray diffraction from single crystal.
[b] The refinement converged to the residual factor ($R$) values.
[c] $j(X)$ – contributions of the intermediate X ion into the AFM ($j(X)<0$) and FM ($j(X)>0$) components of the $Jn$ coupling
[d] $\Delta h(X)$ – the degree of overlapping of the local space between magnetic ions by the intermediate ion X.
[e] $l_n'/l_n$ – asymmetry of position of the intermediate X ion relatively to the middle of the $Cu_i$–$Cu_j$ bond line.
[f] $Cu_iXCu_j$ bonding angle.
[g] $Jn^{str}$ in Å$^{-1}$ and meV – the magnetic couplings ($Jn<0$ - AFM, $Jn>0$ – FM) calculated on the basis of structural.
[h] $Jn^{exp}$ – the exchange interaction parameters extracted from the experimental data (unit: meV) in $Cu_3Mo_2O_9$ (Matsumoto et al. 2012) and $Jn^{str}$ (Å$^{-1}$)×K in $KCu_3OCl(SO_4)_2$. In (Matsumoto et al. 2012) $Jn>0$ – AFM, $Jn<0$ – FM. We changed signs to opposite ones.
[i] $Kn$ – scaling factors ($Kn = Jn^{exp}$ meV/$Jn^{str}$ Å$^{-1}$) for translating the value $Jn^{str}$ in per angstrom into meV.
[j] $j(X^{eg})$ – the contribution of the intermediate X ion occupying equatorial positions in $Cu^{2+}$ octahedra taken into account in calculations of $Jn$ parameters.
[k] $j(X^{ax})$ – the contribution of the intermediate X ion occupying axial positions in $Cu^{2+}$ octahedra not taken into account in calculations of $Jn$ parameters.

The crystal structure of this mineral contains a chain-type oxocentered cationic complex, in which the oxygen atoms O9 are tetrahedrally coordinated by copper atoms Cu1 and Cu2 and two atoms Cu3. These oxocentered octahedra [O9Cu$_4$] are linked through corners (Cu3) into chains stretched along the $c$ axis, while in each individual chain adjacent tetrahedra "face" opposite sides (Figs. 4b, 6). Such one-dimensional fragments can be found in the pyrochlore three-dimensional lattice. The chains of OCu$_4$ tetrahedra in kamchatkite are polarized. Separation of centers of positive and negative charges is expressed in inequality of Cu(3)–O9 bond lengths along the –Cu3–O9–Cu3–O9– chain, in which shortened (to 1.86 Å) and elongated (to 1.98 Å) bonds alternate (Fig. 6).

The latter can be considered as a 0.065 Å shift of whether Cu3 ions in the 001 direction of O9 (tetrahedra-centering) ions in the opposite 00-1 direction. The electric polarization can be eliminated by a 0.065 Å shift of whether Cu3 ions (from the initial value z(Cu3) = 0.0811 to z(Cu3) = 0.0718) or O9 ions (from the initial value z(O9) = 0.8227 to z(O9) = 0.8320) in opposite directions. Finally, both Cu3–O9 lengths along the chain will be 1.92 Å.

According to our calculations (Table 1, Figs. 4b, 5a–f), strong AFM couplings emerge along all edges of the Cu$_4$ tetrahedron. Substantial contributions into AFM components of all these couplings are provided by intermediate oxygen O9 ions centering these tetrahedra. The strongest AFM $J4_1$ ($J4_1^{str}$ = -0.1004 Å$^{-1}$, d(Cu3–Cu3) = 3.501 Å) and AFM $J3$ ($J3^{str}/J4_1^{str}$ = 0.69, d(Cu1–Cu2) = 3.183 Å) couplings are present along two perpendicular tetrahedron edges Cu3–Cu3 and Cu1–Cu2 (Fig. 5e, f).

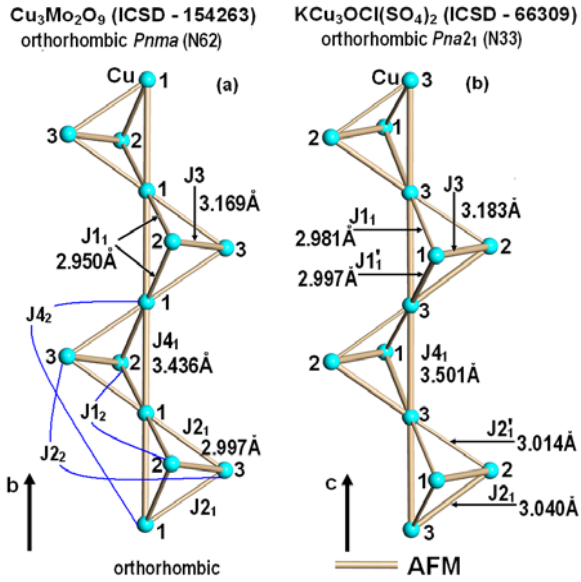

**Fig 4** Single chains of corner-sharing tetrahedra (Cu$_4$) and the coupling $Jn$ in Cu$_3$Mo$_2$O$_9$ (a) and KCu$_3$OCl(SO$_4$)$_2$ (b).

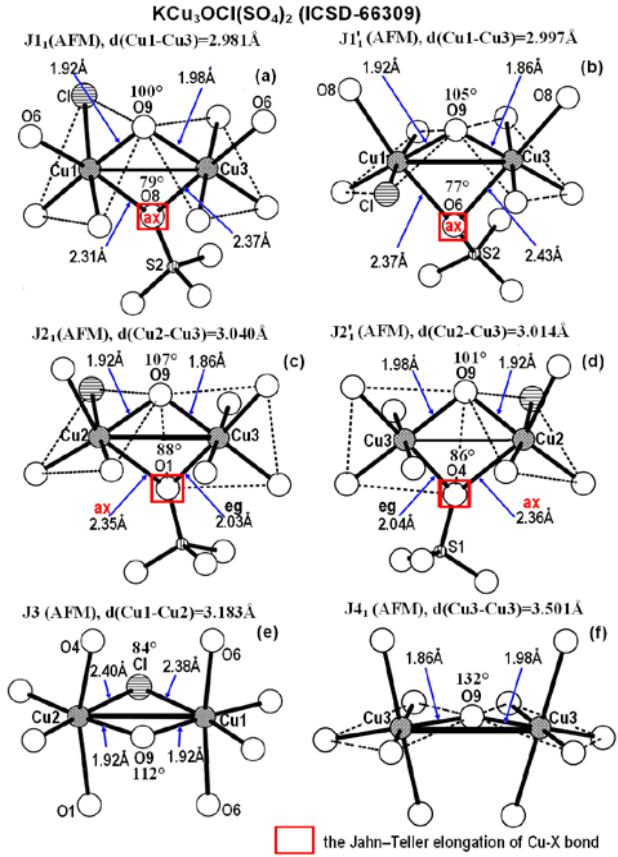

**Fig. 5** The arrangement of intermediate ions in local space of $J1_1$, $J1_1$', $J2_1$, $J2_1$', $J3$ and $J4_1$ couplings in chains of corner-sharing tetrahedra) in KCu$_3$OCl(SO$_4$)$_2$

Rather strong AFM $J1_1$ ($J1_1^{str}/J4_1^{str}$ = 0.32, d(Cu1–Cu3) = 2.981 Å), $J1_1$' ($J1_1$'$^{str}/J4_1^{str}$ = 0.54, d(Cu1–Cu3) = 2.997 Å), $J2_1$ ($J2_1^{str}/J4_1^{str}$ = 0.60, d(Cu2–Cu3) = 3.040 Å) and $J2_1$' ($J2_1$'$^{str}/J4_1^{str}$ = 0.36, d(Cu2–Cu3) = 3.014 Å) couplings are present along other edges of the Cu$_4$ tetrahedron as well (Figs. 4b, 5a–d).

All the couplings in tetrahedra are strongly frustrated (Table 1, Fig. 4, 7). Each of AFM couplings in the tetrahedron compete with four other AFM couplings in $J3$–$J1_1$'–$J2_1$, $J3$–$J1_1$–$J2_1$', $J4_1$–$J2_1$–$J2_1$', and $J4_1$–$J1_1$–$J1_1$' triangles. Besides, there exist extra possibilities for the emergence of competition in pyrochlore chains forming these very AFM tetrahedra (Fig. 4). First, there exists the competition of nearest AFM $J4_1$ couplings between Cu3 ions with next-to-nearest AFM $J4_2$ ($J4_2^{str}/J4_1^{str}$ = 0.30, d(Cu3–Cu3) = 7.001 Å) couplings in linear chains along the $c$ axis. Second, there exists the competition along edges of two corner-sharing tetrahedra between nearest couplings $J1_1$ and $J1_1$' ($J2_1$ and $J2_1$') and next to them $J1_2$ ($J1_2^{str}/J1_1^{str}$ = 1.05 and $J1_2^{str}/J1_1$'$^{str}$ = 0.61, d(Cu1–Cu1) = 5.978 Å) and $J2_2$ ($J2_2^{str}/J2_1^{str}$ = 0.55 and $J2_2^{str}/J2_1$'$^{str}$ = 0.92, d(Cu2–Cu2) = 6.054Å) couplings. There are no other strong intrachain couplings ($J5$ = 0, d(Cu1–Cu3) = 5.743 Å; $J6^{str}$ = 0.0002 Å$^{-1}$ FM, d(Cu2–Cu3) = 5.774 Å; $J7^{str}$ = -0.0006 Å$^{-1}$ AFM, d(Cu1–Cu3) = 5.823 Å; $J8$ = 0, d(Cu2–Cu3) = 5.831 Å). There is no magnetic coupling between Cu1 ions along the $c$ axis ($J_c^{1-1}$ = 0, d(Cu1–Cu1) = 7.001 Å), since the intermediate O8 and O6 ions are axial. There is a weak FM $J_c^{2-2}$ coupling ($J_c^{2-2, str}$ = 0.0093 Å$^{-1}$) between Cu2 ions due to the contribution of the intermediate O3 ion into the ferromagnetic interaction component.

Tetrahedra chains are linked to each other through two strong AFM couplings $J11$ ($J11^{str}$ = -0.0838 Å$^{-1}$, d(Cu1–Cu1) = 5.518 Å) and $J15$ ($J15^{str}$ = -0.0889 Å$^{-1}$, d(Cu1–Cu2) = 7.075 Å) and one threefold weaker AFM $J14$ ($J14^{str}$ = -0.0264 Å$^{-1}$, d(Cu2–Cu2) = 6.333 Å) coupling (Fig. 7a, Table 2). Strong AFM $J11$ (Fig. 7b) and $J15$ (Fig. 7c) couplings emerged under effect of intermediate O2 ions and Cl, respectively. These strong interchain couplings form with AFM $J3$ ones $J11$–$J3$–$J15$ triangles ($J11^{str}/J3^{str}$ = 1.21 and $J15^{str}/J3^{str}$ = 1.28) (Figs. 7a and e), in which they compete with each other and, as a result, increase frustration in tetrahedra chains and between them. Two intermediate oxygen ions participate in formation of AFM $J14$ couplings: O5 and O(7) (Fig. 7e). $J14$ couplings are included into two AFM triangles: $J14$–$J2_1$'–$J13$ ($J14^{str}/J2_1$'$^{str}$ = 0.44 and $J13^{str}/J2_1$'$^{str}$ = 0.05) and $J14$–$J2_1$–$J12$ ($J14^{str}/J2_1^{str}$ = 0.72 and $J13^{str}/J2_1$'$^{str}$ = 0.12). However, the competition in these triangles will be weaker than in the $J11$–$J3$–$J15$ triangle, since the strengths of AFM couplings in them are inequal. Interchain couplings at short distances – FM $J9$ (d(Cu1–Cu2) = 4.973 Å), FM $J10$ (d(Cu1–Cu2) = 4.983 Å), AFM $J12$ (d(Cu2–Cu3) = 6.127 Å), and AFM $J13$ (d(Cu2–Cu3) = 6.174 Å) – are very weak.

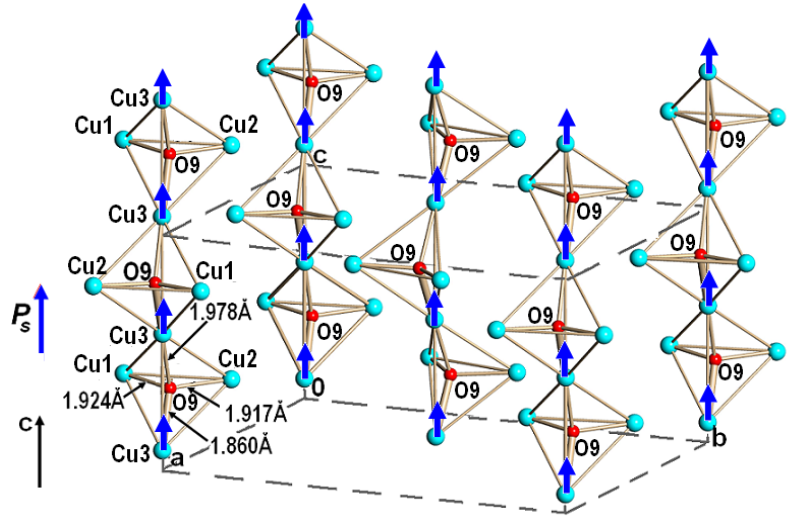

**Fig. 6** Kamchatkite (KCu$_3$O(SO$_4$)$_2$Cl): polarization along the $c$ axis in the 001 directions of [O$_2$Cu$_6$] chains of corner-sharing (OCu$_4$)$^{6+}$ tetrahedra.

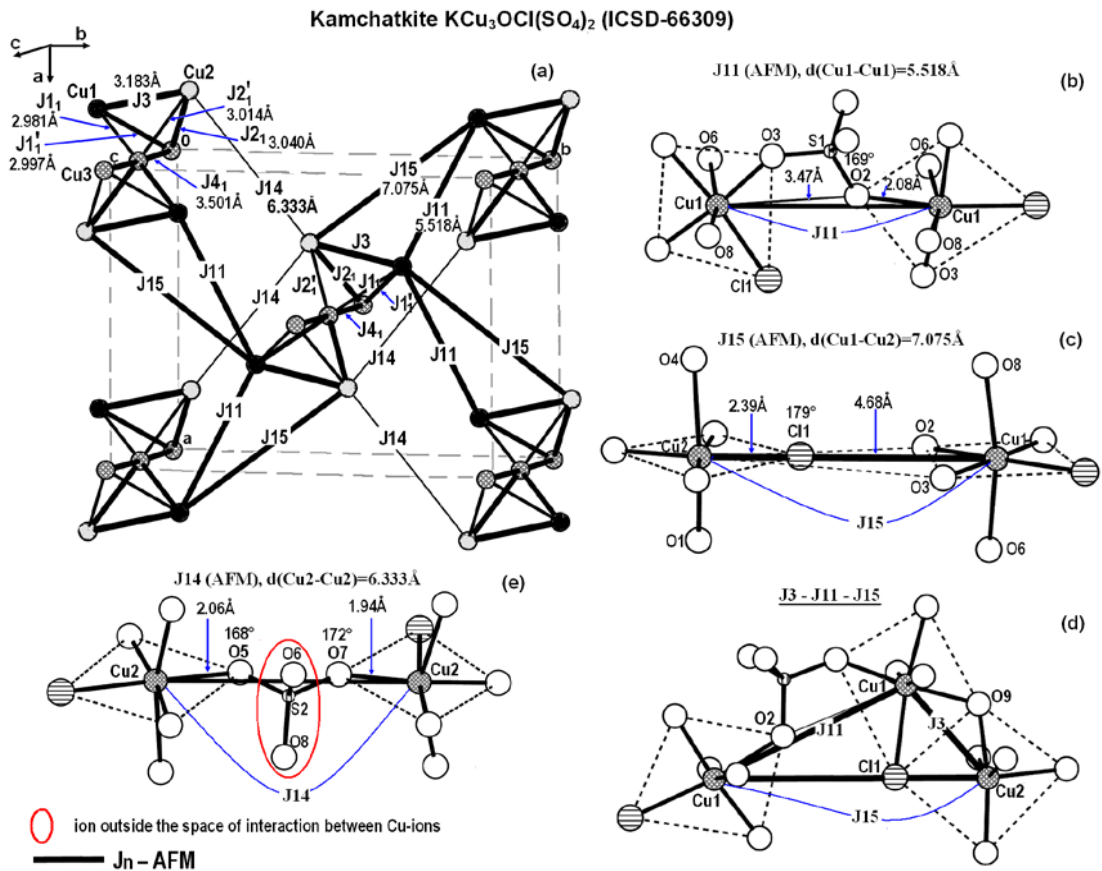

**Fig 7** The view down along [001] intrachain and interchain $J_n$ couplings in KCu$_3$OCl(SO$_4$)$_2$) (the thickness of lines shows the strength of AFM $J_n$ coupling) (a). The arrangement of intermediate ions in local space of interchain $J$11 (b), $J$15 (c), $J$14 (e) couplings and in the frustrated triangle $J$3–$J$11–$J$15 (d).

**Table 2** Parameters of interchain magnetic couplings ($Jn$) calculated on the basis of structural data and respective distances between magnetic ions in kamchatkite $KCu_3OCl(SO_4)_2Cl$

| Crystallographic and magnetic parameters | $KCu_3OCl(SO_4)_2$ (Varaksina et al. 1990) (Data for ICSD - 66309) | |
|---|---|---|
| Bond | Cu1-Cu2 | Cu2-Cu3 |
| d(Cu-Cu) (Å) | 4.973 | 6.174 |
| $j(X)^a$ (Å$^{-1}$) | $j(O3^{eg,(f)})$: -0.0012 | $j(O7^{eg})$: -0.0030 |
| ($\Delta h(X)^b$ Å, $l_n'/l_n^c$, CuXCu$^d$) | (-0.074, 2.56, 116.13°) | (-0.323, 2.84, 132.93°) |
| $j(X)^a$ (Å$^{-1}$) | $j(O7^{eg,(f)})$: 0.0024 | $j(O8^{ax})$: -0.0726 |
| ($\Delta h(X)^b$ Å, $l_n'/l_n^c$, CuXCu$^d$) | (0.224, 3.72, 100.49°) | (-1.240, 1.61, 173.74°) |
| $j(X)^a$ (Å$^{-1}$) | $j(O8^{ax,(g)})$: 0.0056 | $J13 = -0.0030$ [-0.2] |
| ($\Delta h(X)^b$ Å, $l_n'/l_n^c$, CuXCu$^d$) | (0.300, 2.19, 106.06°) | |
| $J^{str(e)}$ (Å$^{-1}$) [meV] | $J9 = 0.0012$ [0.1] | |
| Bond | Cu1-Cu2 | Cu2-Cu2 |
| d(Cu-Cu) (Å) | 4.983 | 6.333 |
| $j(X)^a$ (Å$^{-1}$) | $j(O3^{eg})$: -0.0002 | $j(O5^{eg})$: -0.0132 |
| ($\Delta h(X)^b$ Å, $l_n'/l_n^c$, CuXCu$^d$) | (-0.012, 2.74, 113.03°) | (-1.107, 2.10 167.93°) |
| $j(X)^a$ (Å$^{-1}$) | $j(O7^{eg})$: 0.0015 | $j(O7^{eg})$: -0.0132 |
| ($\Delta h(X)^b$ Å, $l_n'/l_n^c$, CuXCu$^d$) | (0.115, 3.13, 106.67°) | (-1.208, 2.29, 171.80°) |
| $j(X)^a$ (Å$^{-1}$) | $j(O6^{ax})$: 0.0285 | $J14 = -0.0264$ [-2.1] |
| ($\Delta h(X)^b$ Å, $l_n'/l_n^c$, CuXCu$^d$) | (0.285, 1.98, 107.82°) | |
| $J^{str(e)}$ (Å$^{-1}$) [meV] | $J10 = 0.0013$ [0.1] | |
| Bond | Cu1-Cu1 | Cu1-Cu2 |
| d(Cu-Cu) (Å) | 5.518 | 7.075 |
| $j(X)^a$ (Å$^{-1}$) | $j(O2^{eg})$: -0.0856 | $j(Cl^{eg})$: -0.0874 |
| $\Delta h(X)^b$ Å, $l_n'/l_n^c$, CuXCu$^d$) | (-1.147, 1.68, 168.81°) | (-1.773, 1.95, 178.67°) |
| $J^{str(e)}$ (Å$^{-1}$) [meV] | $J11 = -0.0838$ [-6.5] | $J15 = -0.0889$ [-6.9] |
| Bond | Cu2-Cu3 | |
| d(Cu-Cu) (Å) | 6.127 | |
| $j(X)^a$ (Å$^{-1}$) | $j(O7^{eg})$: -0.0043 | |
| ($\Delta h(X)^b$ Å, $l_n'/l_n^c$, CuXCu$^d$) | (-0.426, 2.66, 137.47) | |
| $j(X)^a$ (Å$^{-1}$) | $j(O6^{ax})$: -0.0690 | |
| ($\Delta h(X)^b$ Å, $l_n'/l_n^c$, CuXCu$^d$) | (-1.185, 1.53, 171.62°) | |
| $J^{str(e)}$ (Å$^{-1}$) [meV] | $J12 = -0.0043$ [-0.3] | |

[a] $j(X)$ – contributions of the intermediate X ion into the AFM ($j(X) <0$) and FM ($j(X)>0$) components of the $Jn$ coupling
[b] $\Delta h(X)$ – the degree of overlapping of the local space between magnetic ions by the intermediate ion X.
[c] $l_n'/l_n$ – asymmetry of position of the intermediate X ion relatively to the middle of the $Cu_i$–$Cu_j$ bond line.
[d] $Cu_iXCu_j$ bonding angle.
[e] $Jn^{str}$ in Å$^{-1}$ and in meV ($Jn$ (meV) = $Jn$ (Å$^{-1}$)×K, where scaling factors $K_{middle} = 78$) – the magnetic couplings ($Jn<0$ - AFM, $Jn>0$ – FM).
[f] $j(X^{eg})$ – the contribution of the intermediate X ion occupying equatorial positions in $Cu^{2+}$ octahedra taken into account in calculations of $Jn$ parameters.
[g] $j(X^{ax})$ – the contribution of the intermediate X ion occupying axial positions in $Cu^{2+}$ octahedra not taken into account in calculations of $Jn$ parameters

Unlike kamchatkite ($KCu_3O(SO_4)_2Cl$), we have not found strong couplings between spin-frustrated pyrochlore chains in $Cu_3Mo_2O_9$. The latter comprises a basic difference of the magnetic structures under examination.

Let us compare the parameters of magnetic couplings $Jn^{str}$ for $KCu_3OCl(SO_4)_2$ and $Cu_3Mo_2O_9$ (Table 1) with the experimental data $Jn^{exp}$ for $Cu_3Mo_2O_9$ (Matsumoto et al. 2012). In spite of the determining role of structural factors in the magnetic lattice formation, there exist other factors contributing to this process. Besides, the impossibility of direct estimation, using the crystal chemistry method, of the effect of the magnetic couplings competition in the magnetic lattice on their strength and insignificant deviations of the composition and crystal structure of real crystals from the data on their ideal structure could result, in some cases, in significant differences with the experiment.

The values of $J1_1^{str}$ and $J2_1^{str}$ we calculated for $Cu_3Mo_2O_9$ (Table 1, Fig. 4a) differ ($J1_1^{str}/J2_1^{str} = 1.34$), although they both $J1_1^{exp}$ and $J2_1^{exp}$ amount to -3.06 meV in (Matsumoto et al. 2012). The main reason of the difference between the $J1_1^{str}$ and $J2_1^{str}$ values consists in the fact that, aside from the intermediate O1 ion present in the local space of interaction of both couplings and making substantial AFM $j(O1)$ contributions of -0,0469 Å$^{-1}$ and -0.0503 Å$^{-1}$ into $J1_1$ and $J2_1$, respectively, their local spaces each contain one more oxygen atom. In addition, the local space of the $J1_1$ coupling contains the O5 ion having an axial bond with the Cu1 ion and, therefore, its FM contribution (j(O5$^{ax}$) = 0.0267) is not taken into account (see above). The local space of the $J2_1$ coupling additionally contains the O4 ion having just equatorial bonds with copper ions, makes an FM contribution (j(O4$^{eg}$) = 0.0152), and, as a result, decreases the strength of the AFM $J2_1^{str}$ coupling down to -0.0351 Å$^{-1}$. It is worth mentioning that this O4 ion is localized near the critical "b" position (h(O4$^{eg}$) ≈ r(O$^{2-}$)) (Volkova and Polyshchuk 2005, 2009), so that in the case of its shift by 0.067 Å to the Cu1–Cu3 bond line its FM contribution to the $J2_1$ interaction disappears. As a result, $J2_1^{str}$ becomes equal to the AFM contribution only from the O1 ion (j(O1) = -0.0503 Å$^{-1}$), and the $J1_1^{str}$ and $J2_1^{str}$ values become virtually equal ($J1_1^{str}/J2_1^{str}$ = 0.93).

In KCu$_3$OCl(SO$_4$)$_2$, the Cu1–Cu2 bond of Cu$_3$Mo$_2$O$_9$ is turned into two in equivalent Cu1–Cu3 bonds (Table 1, Fig. 4). At the first glance, two Cu1–Cu3 bonds have relatively similar lengths and surroundings, but the extracted coupling coefficients of $J1_1^{str}$ and $J1_1'^{str}$ are quite different (-0.0318 vs -0.0547 Å$^{-1}$). Such a difference emerges because of the inequality of AFM j(O9) contributions made by the intermediate O9 ions to the formation of $J1_1$ (j(O9) = -0.0318 Å$^{-1}$) and $J1_1'$ (j(O9) = -0.0547 Å$^{-1}$) couplings. The point is, the O9 ion is located much closer to the line of the Cu1–Cu3 bond in $J1_1'$ (Δh(O9) = -0.245 Å, Cu1O9Cu3 = 104.75°) than in $J1_1$ (Δh(O9) = -0.141 Å, Cu1O9Cu3 = 99.61°) (Figs. 2b, 5ab, Table 1). The same factor is responsible for the emergence of differences between the $J2_1^{str}$ and $J2_1'^{str}$ couplings (Figs. 5cd, Table 1). In the course of the interaction of magnetic ions located at short distances, even insignificant shifts of intermediate ions in the local interaction space induce significant changes in the forces of magnetic couplings. Most probably, the experimental methods are not so sensitive to local changes in the forces of magnetic couplings or the determination of the crystal structure and the magnetic couplings parameters was carried out at different temperatures.

The effect of frustration on the strength of magnetic couplings can be demonstrated on the examples below. The $J3^{str}$ and $J4_1^{str}$ values we calculated (Fig. 5, Table 1) without taking into account the frustration for Cu$_3$Mo$_2$O$_9$ differ more significantly ($J3^{str}/J4_1^{str}$ = 0.47) than their experimental values ($J3^{exp}/J4_1^{exp}$ = 0.88) determined in (Matsumoto et al. 2012). This must be related to the fact that the real $J4_1^{exp}$ value in Cu$_3$Mo$_2$O$_9$ decreases more significantly in comparison with the $J3^{exp}$ one because of the presence of an additional competition of the nearest AFM $J4_1$ couplings with the next-to-nearest AFM $J4_2$ couplings in linear chains along the c axis. The $J3$ couplings compete only with those along the triangular tetrahedron faces, unlike the $J4_1$, $J1_1$, and $J2_1$ couplings having extra competition with the AFM $J4_2$, $J1_2$, and $J2_2$ couplings, respectively.

Extra competition characterizes the $J3$ couplings of kamchatkite (KCu$_3$OCl(SO$_4$)$_2$) with strong interchain AFM $J11$ and $J15$ couplings in $J3$–$J11$–$J15$ triangles. The presence of this extra competition must strongly decrease the strength of the $J3$ couplings on KCu$_3$O(SO$_4$)$_2$Cl, unlike Cu$_3$Mo$_2$O$_9$, in which the interchain bonds are very weak. Because of the differences in the degree of frustration of the $J3$ couplings in Cu$_3$Mo$_2$O$_9$ and KCu$_3$O(SO$_4$)$_2$Cl, the use of the scaling factor $K3$ calculated from the Cu$_3$Mo$_2$O$_9$ data will yield the overstatement of the $J3$ value. Indeed, in KCu$_3$O(SO$_4$)$_2$Cl the value of the ratio $J3^{str}/J4^{str}$ = 0.69 determined on the basis of the structural data is substantially smaller than the value $K3 \times J3^{str}/K4 \times J4^{str}$ = 1.29 calculated using the scaling factor $K3$.

As was shown above, the studies of the magnetism of Cu$_3$Mo$_2$O$_9$ (unlike KCu$_3$OCl(SO$_4$)$_2$) have been described in numerous works. In particular, it was established [Hamasuki et al. 2008] that an AF second-order phase transition was observed at $T_N$ = 7.9 K and the weak ferromagnetic phase transition occurred at $T_c$ = 2.5 K at zero magnetic field. These results are explainable by both magnetic frustration among symmetric exchange interactions and competition between symmetric and asymmetric Dzyaloshinskii–Moriya ones. Our attempts to find data on magnetic properties of kamchatkite were not successful. However, as was reported just recently, the authors of two works [Kikuchi et al. 2017 and Kunieda, Kikuchi et al. 2016] revealed the existence of several low-temperature magnetic transitions (3, 11 and 15 K) as well as at Curie Weiss temperatures (-21 K), with participation of spin frustration. Unfortunately, using our crystal chemistry method, we are not capable to estimate the temperatures of phase transitions $T_c$ and $T_N$, but, nevertheless, we can identify structural factors, which could participate and even be responsible for the emergence of such transitions.

According to our calculations of parameters of magnetic couplings, a common factor for the emergence of all these transitions in Cu$_3$Mo$_2$O$_9$ and KCu$_3$OCl(SO$_4$)$_2$ is, undoubtedly, the geometric frustration considered above in detail. In order to determine whether magnetic transitions are related to changes in the crystal structure of Cu$_3$Mo$_2$O$_9$ upon the temperature decrease, we additionally calculated the parameters of magnetic couplings in this compound at a temperature of 2 K (ICSD – 173770, Vilminot et al. 2009), which is by 0.5 K lower than that of the phase transition, and compared them with respective values at room temperature. It turned out that all the parameters except one ($J2_1^{str}$) (Fig.5) were virtually identical at both temperatures. The value of the $J2_1^{str}$ parameter increased up to -0.0430 Å$^{-1}$ and became virtually identical to that of $J1_1^{str}$ ($J1_1^{str}/J2_1^{str}$ = 1.1), as was determined experimentally (Table 1) in (Matsumoto et al. 2012). The latter was the result of the decrease of the FM contribution j(O4$^{eg}$) down to 0.063 Å$^{-1}$ at an insignificant shift of the intermediate O4 ion (Δh(O4) = 0.028 Å, l'/l = 1.12, Cu1O4Cu3 = 92.33°) to the Cu1–Cu3 bond line upon the temperature decrease.

To study the effect of the loss of the inversion center and the emergence of polarization because of the shift of O1 ions from the centers of Cu4 tetrahedra along the *c* axis in the -001 direction, we used the crystal structure of $Cu_3Mo_2O_9$, which was initially described in the noncentrosymmetric space group *Pna*$2_1$ (Kihborg et al. 1972). Since the centrosymmetric-noncentrosymmetric modification transition is accompanied with insignificant atomic shifts, this enables one to assume the possibility of realization of transitions of this type and, as a result, coming into effect of Dzyaloshinskii-Moriya (DM) forces. In spite of a negligibility of changes in the crystal structure at such a transition, the magnetic structure undergoes important alterations. The results of calculations of parameters of magnetic couplings in the noncentrosymmetric model demonstrate that a significant anisotropy (difference) of magnetic coupling forces varying within broad limits (-0.0101 Å – -0.0611 Å) additionally emerges in each pair composed of couplings split into two: $J1_1$ ($J1_1'^{str}/J1_1^{str}$ = 0.18) and $J2_1$ ($J2_1'^{str}/J2_1^{str}$ = 0.32). Here, the parameters of perpendicular to each other magnetic couplings in *J*3 and *J*4 undergo virtually no changes.

Finally, we calculated the course of changes of the parameters of magnetic couplings in $KCu_3OCl(SO_4)_2$, if one eliminates, within the frames of the noncentrosymmetric space group *Pna*$2_1$, the electric polarization in tetrahedral chains through shifting (by 0.065 Å along the (00-1) direction) of O9 ions into the tetrahedra centers. As was expected, such a procedure virtually equalized the forces of four magnetic couplings ($J1_1^{str}$, $J1_1'^{str}$, $J2_1^{str}$, and $J2_1'^{str}$), whose values became varied within narrow limits (-0.0426 Å – -0.0501 Å), unlike the polarized structure (-0.0318 Å – -0.0605 Å) (Table 1). Also, the above shift does not virtually affect the values of $J3^{str}$ and $J4^{str}$ couplings.

To sum up, the temperature decrease, as well as the elimination of the electric polarization in the tetrahedral chain (the shift of the centering atom to the tetrahedron center) decrease the anisotropy of forces of the $J1_1^{str}$, $J1_1'^{str}$, $J2_1^{str}$, and $J2_1'^{str}$ magnetic couplings and, thus, increase the frustration in tetrahedra. In opposite, the emergence of polarization and the elimination of the inversion center increase the anisotropy of the above forces, which, along with some decrease of their competition and the loss of the inversion center, promotes the emergence of the DM interaction.

## Conclusions

We have determined the parameters (sign and strength) of magnetic couplings in kamchatkite ($KCu_3OCl(SO_4)_2$) based on structural data. As shown by the calculation results, the kamchatkite magnetic system contains AFM spin-frustrated pyrochlore chains composed of corner-sharing Cu4 tetrahedra. Competition in AFM chains exists not only between nearest couplings along tetrahedra triangular faces, but also between nearest and next-to-nearest neighbors inside the chain. Besides, there exists the interchain competition in triangles composed of strong AFM intrachain and interchain couplings. In opposite to frustration of the spin structure, the electric polarization along the *c* axis exists in klyuchevskite.

Magnetic frustration emerges in the ordered crystal structure thanks to geometric considerations. Oxocentered copper OCu4 tetrahedra forming the basis the crystal structure serve as a platform for geometric frustration of the magnetic system of not only kamchatkite, but also many minerals of Kamchatka Tolbachik volcanos. As we demonstrated in [Volkova and Marinin, 2017], the uniqueness of these systems consisted in the fact that the antiferromagnetic character of couplings along the tetrahedron edges and, therefore, frustration of exchange interactions on triangular faces were caused mainly by oxygen ions centering copper tetrahedra. This oxygen ion is an intermediate one simultaneously in all six couplings along the tetrahedron edges and makes a substantial contribution to formation of the AFM character of these couplings. Reorientation of magnetic moments (AFM → FM) along the tetrahedra edges and, as a result, suppression of frustration due to changes in the character of exchange interactions in them will be rather complicated, since displacement of these oxygen ions is limited by small sizes of Cu4 tetrahedra. It is assumed that frustration of the magnetic system in kamchatkite ($KCu_3OCl(SO_4)_2$) can fully block the formation of the long-range order until realization of exotic states of "spin ice" or "spin liquid" types.

**Acknowledgments** The work was partially supported by the Program of Basic Research "Far East" (Far-Eastern Branch of Russian Academy of Sciences), project no. 15–I–3–026.